\documentclass[aip,
 rsi,
 amsmath,amssymb,
 reprint,%
floatfix,
]{revtex4-1}

\usepackage{graphicx}
\usepackage{dcolumn}
\usepackage{bm}
\usepackage{soul}
\usepackage{xcolor}
\usepackage[utf8]{inputenc}
\usepackage[T1]{fontenc}
\usepackage{mathptmx}
\usepackage{etoolbox}
\usepackage{float}
\graphicspath{{Figs/}}
\usepackage{xfrac}

\makeatletter
\def\@email#1#2{%
 \endgroup
 \patchcmd{\titleblock@produce}
  {\frontmatter@RRAPformat}
  {\frontmatter@RRAPformat{\produce@RRAP{*#1\href{mailto:#2}{#2}}}\frontmatter@RRAPformat}
  {}{}
}%
\makeatother

\begin{document}

\preprint{AIP/123-QED}

\title[Cryogenic probe for low-noise, high-frequency electronic measurements]{Cryogenic probe for low-noise, high-frequency electronic measurements}
\author{E. Garcia}
 \affiliation{Department of Physics, Brown University, Providence, RI 02912, USA}
\author{C. Bales}%
\affiliation{Department of Physics, Brown University, Providence, RI 02912, USA}%
\author{W. Patterson}%
\affiliation{School of Engineering, Brown University, Providence, RI 02912, USA}%
\author{A. Zaslavsky}%
\affiliation{Department of Physics, Brown University, Providence, RI 02912, USA}%
\affiliation{School of Engineering, Brown University, Providence, RI 02912, USA}%
\author{V. F.  Mitrovi\'{c}}
  \homepage{Author to whom correspondences should be addressed: vemi@brown.edu}
\affiliation{Department of Physics, Brown University, Providence, RI 02912, USA}%
\affiliation{School of Engineering, Brown University, Providence, RI 02912, USA}%

\date{\today}

\begin{abstract}
The design and performance of a low-noise, modular cryogenic probe, applicable to a wide range of measurements over a broad range of working frequencies, temperatures, and magnetic fields is presented. The design of the probe facilitates the exchange of sample holders and sample-stage amplifiers, which, combined with its characteristic low transmission and reflection loss, make this design suitable for high precision or low sensitivity measurements. The specific example of measuring the shot noise of magnetic tunnel junctions is discussed. We highlight various design characteristics chosen specifically to expand the applicability of the probe to measurement techniques such as nuclear magnetic resonance (NMR). 
\end{abstract}

\maketitle

\section{Introduction}

Measurements of electronic noise, where noise fluctuations can be a source of information that is not present in time-averaged values, has long been an informative tool that provides insight into charge transport properties in mesoscopic systems. For example, shot noise has been used to study the transport of fractional charges in the fractional quantum Hall regime\cite{shot_de1998direct,shot_saminadayar1997} and to study the transport of Cooper pairs in superconductor-normal junctions\cite{jehl2000detection}. Thus, the natural question arises as to whether analogous measurements can be performed for evidence of fractionalization of spin degrees of freedom. To implement such measurements, one requires a sensor that can discern ultra-weak magnetic fields with both high temporal and spatial resolutions. 

Sensors based on magnetic tunnel junctions (MTJs), where coherent quantum tunneling occurs through a thin insulating barrier between spin-polarized ferromagnetic electrodes \cite{GangBook11,he2018picotesla,PhysRevB.83.144416}, can, in principle, quantify magnetic fields with high temporal and spatial resolution simultaneously.
Recently, MTJ sensors have demonstrated increased sensitivity \cite{he2018picotesla} and offer much flexibility in their functional parameter space, including a wide range of operating temperatures and a broad frequency response, possibly into the GHz range. MTJs can potentially probe the nature of spin currents carried through edge states in topological insulators by using arrays of such sensors that simultaneously measure correlations between local response functions at different locations in a sample \cite{PhysRevB.104.045144}. Furthermore, MTJs can be employed to investigate the nature of the emergent quantum phases of matter \cite{PhysRevB.104.L060403} by measuring spin current fluctuations.  However, such measurements demand a push to higher frequencies, well beyond the characteristic $1/f$ noise of MTJs, where shot noise can be observed and effectively measured, all while offering extremely low losses and preserving the characteristic impedance of the signal line. 

\begin{figure*}[t]
\includegraphics[scale=0.42]{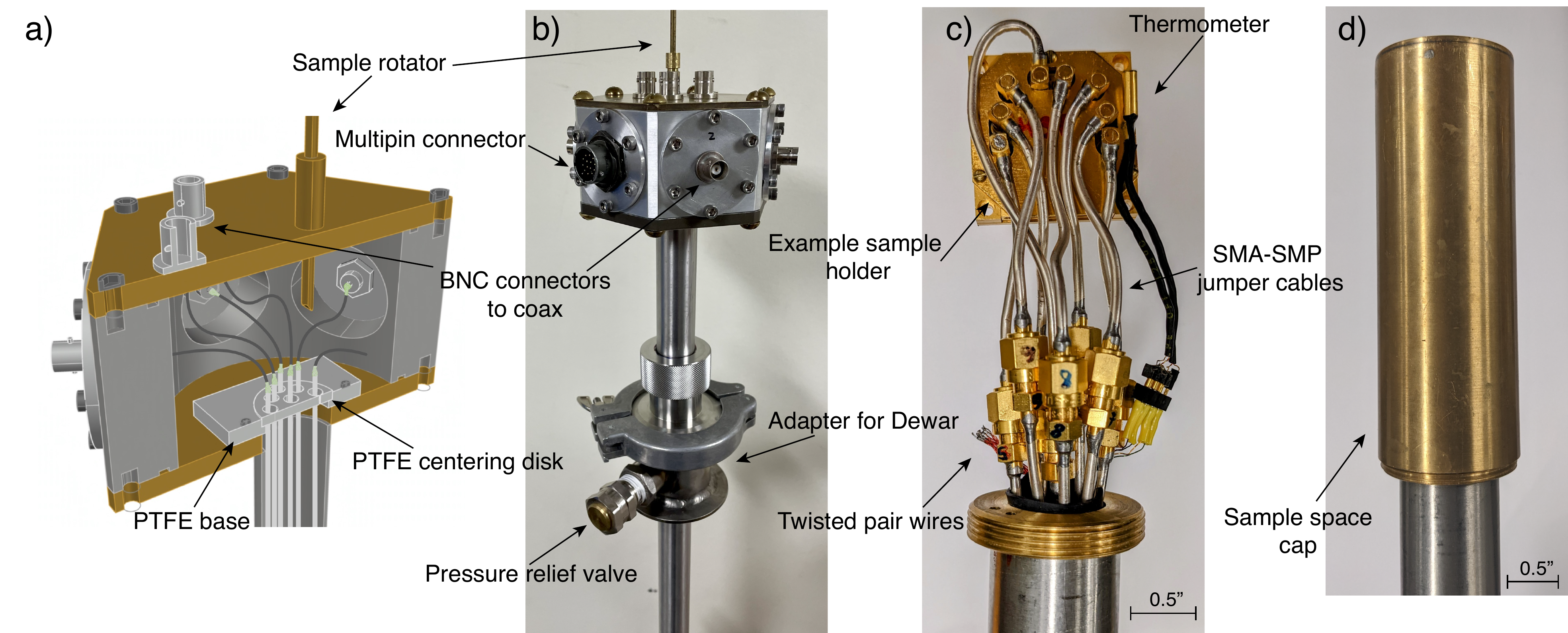} 
\caption{\label{fig_probe} (a) Schematic cut-away view of the inside of the probe head. (b) Picture of the probe head displaying multiple coaxial connections and a multi-pin hermetic connector. (c) Foot of the probe showing a sample holder hooked up with semi-rigid coaxial SMA-SMP jumpers. (d) Sample space cap threaded onto the probe over sample holder and jumper cables.}
\end{figure*}

MTJs present an alternative and complementary method to presently used quantum magnetic sensing technology such as semiconductor-based Hall effect sensors \cite{Vervaeke09} (compatible with silicon technology, but far less sensitive) and diamond nitrogen-vacancy (NV) center magnetometry \cite{PhysRevB.105.024507, PhysRevX.5.041001} (very sensitive, but without placement control and requiring optical readout).
MTJ sensors forgo the need for bulky optical and mechanical components, requiring only a low-noise, low-loss probe. In addition, MTJs have micron scale spatial resolution that grants them a much broader range of applications, such as probing the effects of topological entanglement in quantum materials and for dark matter field detection up to the GHz frequency range \cite{Budker14,Kimball:2017vd,Abel17,Graham18,Alexander18}.

The pursuit of high-sensitivity, high-frequency measurements motivates the need for a probing instrument that is able to perform within the required bandwidth while experiencing a low loss of signal throughout the entire working frequency range. In addition, the probing instrument should also possess low intrinsic noise levels and should be able to mitigate the noise introduced by external instruments connected to the device under test. The measurements are further complicated by the cryogenic temperatures required to observe the exotic spin physics of interest that must be considered during the construction of the instrument. Many flexible and ingenious probe designs for NMR, EPR, and noise measurements have been previously described\cite{reyes1997versatile,annino1999novel,tewari2017fast,parmentier2011high}, but these are designed with some limitations either in signal loss, bandwidth, temperature, or other relevant parameters.

Sophisticated dipsticks have previously been constructed and implemented in shot noise measurements of atomic and molecular junctions\cite{tewari2017fast, tewari2018anomalous}. Here we present a simple but effective alternative optimized for MTJ sensing applications. We discuss the design and performance of a low-loss and high frequency probe suited for cryogenic ultra-sensitive measurements, such as characterizing shot noise and spin fluctuations and detecting NMR-like signals using MTJ devices. The probe performance was tested up to 1.5 GHz and has a -3dB signal transmission loss point of 500 MHz through a single coaxial line of the probe, all while efficiently preserving a \hbox{$50\, \Omega$} impedance, pushing the sensitivity limits of our experimental set-up to those set by the input noise of external instrumentation. In addition, the modular nature of the probe allows for low-temperature amplifiers and custom sample holders for other types of measurements to be added with ease.

\section{\label{sec:level1}Instrument Design}

The general design of the probe is shown in Fig. \ref{fig_probe}. The probe was designed for insertion in standard commercial liquid helium Dewars but can be easily adapted for a custom cryostat by an external coupler slid over the body of the probe to make a vacuum seal, as shown in Fig. \ref{fig_probe}(b). The probe head is constructed from aluminum and measures 2.5 inches in height. The cross sectional view of the probe head can be seen in Fig. \ref{fig_probe}(a). The probe head allows for the easy addition of pumping ports, relief valves, pressure gauges, and gas injection lines when working with closed-cycle cryostat systems and for the easy interchange of connectors. An LC resonant circuit at the room temperature stage or at the foot of the probe can be added to adapt the probe for NMR or microwave measurements. 

\begin{figure}[b!]
\includegraphics[scale=0.37]{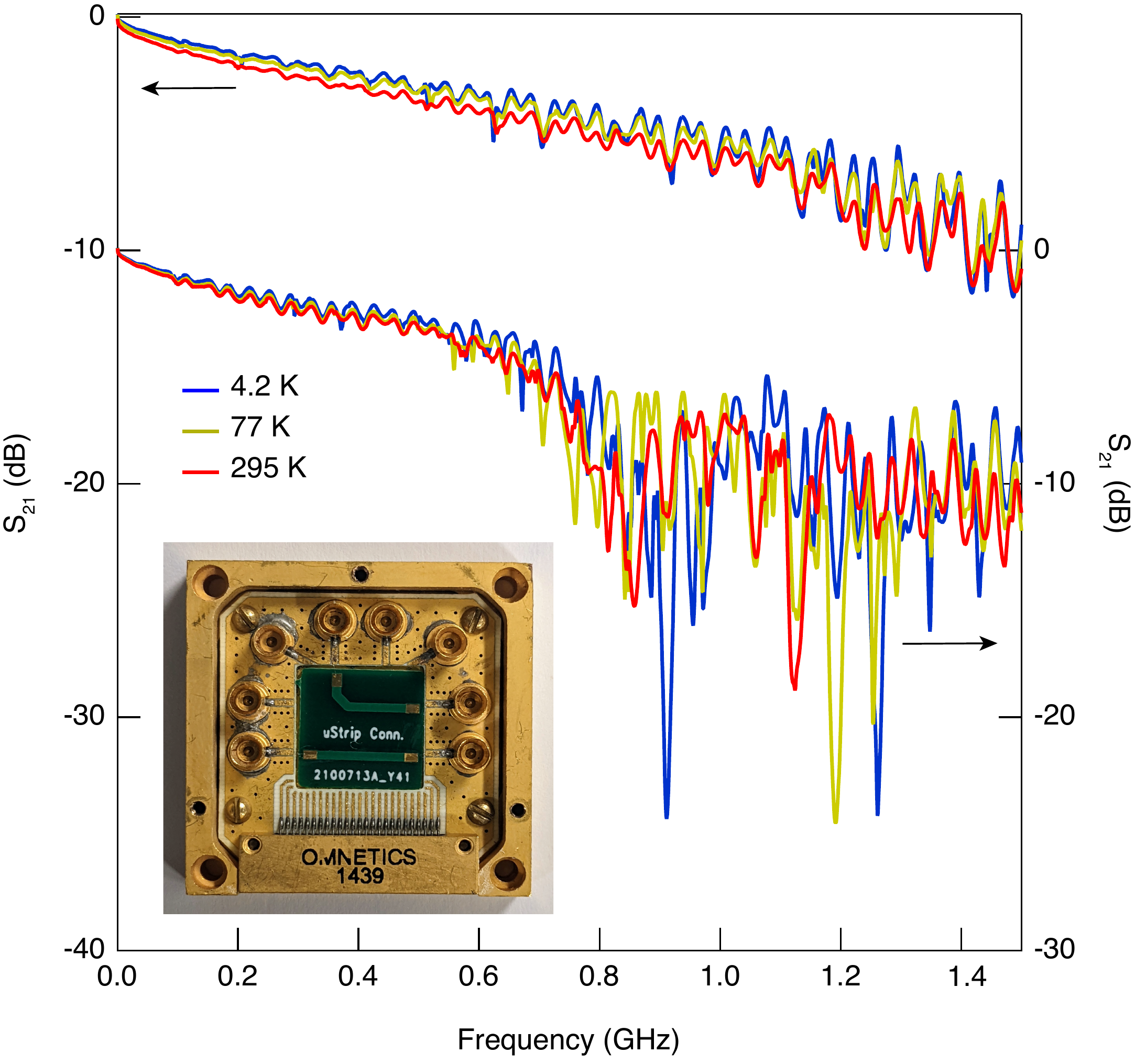}
\caption{\label{fig_s21} Representative coaxial S$_{21}$ measurement, a measurement of the transmission loss, at various temperatures. Four coaxial lines are connected pairwise by two \hbox{$50\, \Omega$} microstrip lines (one straight, the other bent) using wire bonds where the microstrip lines abut the coaxial connectors. The transmission through the straight microstrip line is shown plotted against the left axis and the transmission through the bent line against the right axis. Inset shows the sample holder with the wire-bonded microstrips.}
\end{figure}

The probe tube body is type 304 stainless steel tubing, 53 inches long with a 1.00 inch outer diameter and 0.049 inch wall thickness, that can hold steep thermal gradients. The probe has eight semi-rigid micro-coaxial lines running through the body (type UT-085C-SP). The coaxial lines' outer and inner conductors are made from silver plated copper for optimal conductivity. In the head of the probe, the coax is broken and connected to hermetic \hbox{$50\, \Omega$} BNC connectors by a short, flexible coax line (type RG-174/U). The outer shielding of the flexible coax line is soldered to the outer shielding of the semi-rigid coax line and to that of the BNC connector to maintain a common ground. The coaxial lines are guided through the body of the probe by PTFE centering rings that are suspended by a 304 stainless steel rod that provides vertical support. The coax go through another centering ring at the head of the probe that sits inside a carved-out PTFE base, shown in Fig. \ref{fig_probe}(a). This allows for both the coaxial lines and centering rings to rotate freely throughout their entire length without placing any strain on individual soldering joints at the head of the probe.

To allow precise sample temperature monitoring and control, the probe has a Cernox thin film resistance thermometer (Lake Shore Cryogenics no. CX-1010-AA) at the base of the probe, shown in Fig. \ref{fig_probe}(c), connected by twisted pair wire (Lake Shore Cryogenics no. WQT-32-100) to a multi pin hermetic connector. The probe also has 6 additional twisted pair wires, a total of 12 wires, for DC connections that are accessed through another multi-pin hermetic connector (Amphenol no. PT07A-12-14P).

The base of the probe, shown in Fig. \ref{fig_probe}(c) and (d), is protected by a threaded brass cap that is 4 inches long and has a $1\sfrac{1}{3}$ inch inner diameter that fits over the sample holder and jumper cables and threads onto the body of the probe. As seen in Fig. \ref{fig_probe}(c), the coaxial lines terminate in SMA connectors that are connected to our sample holder by smaller coaxial jumpers for easy adaptability to various sample holders and set-ups. The jumpers are made from semi-rigid micro-coax lines with SMA connectors on one end and mini SMP connectors on the other that plug directly onto the sample holder. The jumpers contribute to the overall modular nature of the probe as they can easily be changed to fit other sample holders with different connectors. 

The probe was designed for operation at both high and low magnetic field environments. The outer body and all internal components are made from nonmagnetic materials. Intended applications included magnetic resonance at several Tesla and MTJ-based magnetic field sensing and noise measurements at fields below 1 G. For experiments with the most stringent requirements for low fields, a sleeve of magnetic shielding material can be slipped into the sample space cap to further reduce the local parasitic field.

\section{Performance}

To characterize the performance of the probe at a wide temperature (295 K - 4.2 K) and frequency (300 kHz - 1.5 GHz) range, we used an Agilent E5061A vector network analyzer (VNA) for S-parameter measurements. Typical results of S$_{21}$ measurements, a measure of the power transferred from port 1 to port 2, at 295 K, 77 K, and 4.2 K for the probe coaxial lines are plotted in Fig. \ref{fig_s21}. The loss displayed in Fig. \ref{fig_s21} is for a signal being transmitted into the probe through one of the coaxial lines and out the other. The two lines are connected by a microstrip with \hbox{$50\, \Omega$} impedance that is wire bonded to the sample holder, shown in the inset of Fig. \ref{fig_s21}, which is connected to the coaxial lines through the jumper cables, as displayed in Fig. \ref{fig_probe}(c). The -3dB point of the transmitted signal is lowest at room temperature at around 400 MHz, and highest at 4.2 K in the vicinity of 550 MHz. Omitting the sample holder completely and measuring S$_{21}$ by shorting two coax lines with an SMA-SMA jumper cable pushed the -3dB point of the transmission to above 1 GHz, which provides insights into the transmission loss incurred by the sample holder and indicates that a different sample holder utilizing different connections to the sample may yield higher transmission at high frequencies. Identical S$_{21}$ measurements on twisted pair wire showed a -3dB point of the transmitted signal below 10 MHz, confirming the obvious benefits of coaxial wiring. 

\begin{figure}[t!]
\includegraphics[scale=0.42]{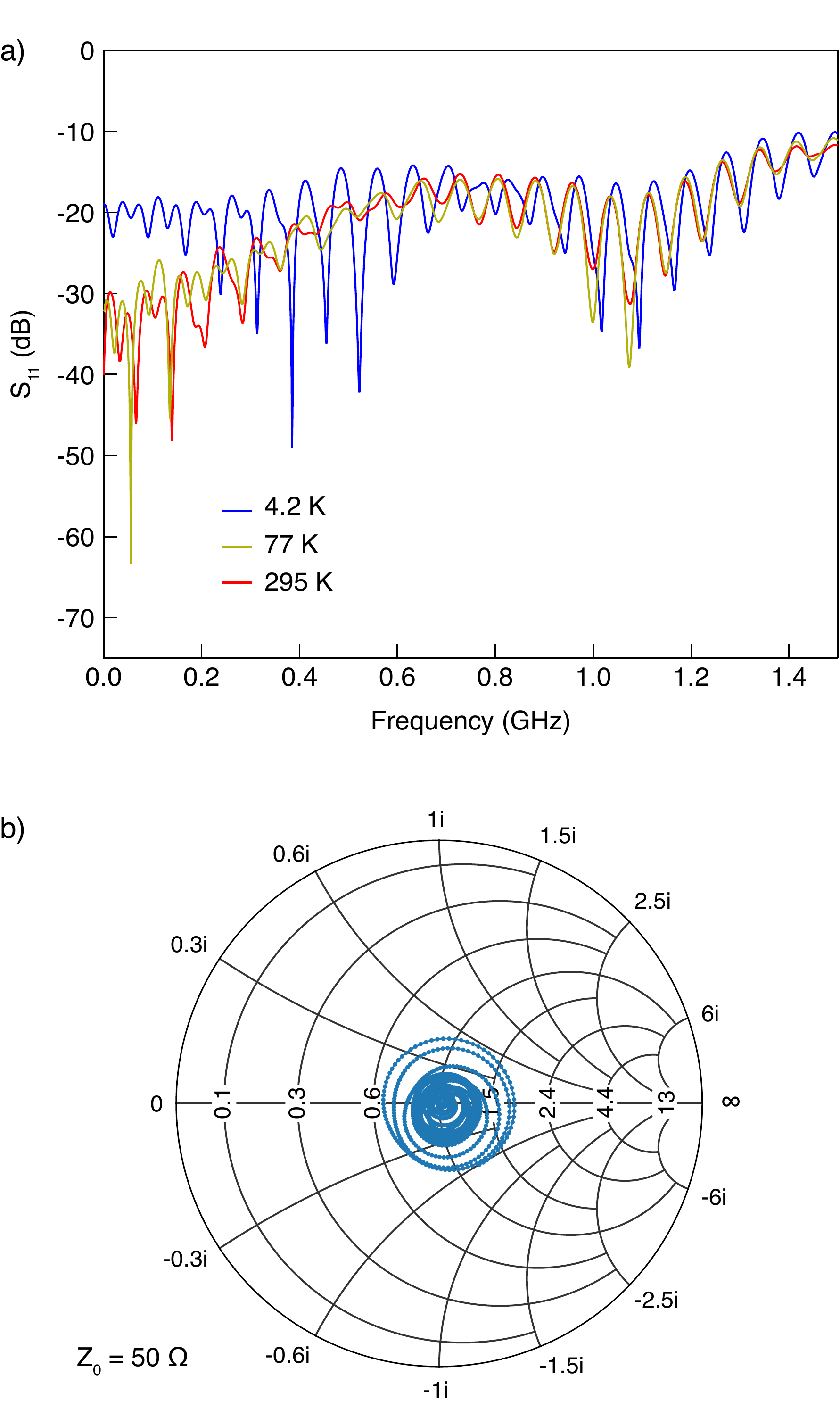}
\caption{\label{fig_s11} (a) Representative coaxial S$_{11}$ measurement, a measure of the return loss, at various temperatures. (b) Smith chart for corresponding 77 K measurement of (a).}
\end{figure}

S$_{11}$ measurements, a measure of the power reflected back to the same port it is transmitted from, were also performed to characterize the return loss of our probe. A representative S$_{11}$ measurement for our coaxial lines is illustrated in Fig. \ref{fig_s11}(a). The Smith chart shown in Fig. \ref{fig_s11}(b) demonstrates the degree to which the impedance is matched within our probe. Much effort was invested into preserving the \hbox{$50\, \Omega$} impedance of the coaxial lines at the joints between room temperature and low temperature lines. Wrapping the inner conductor junction with PTFE and wrapping the joint with the braided outer conductor of the RG-174/U coaxial cable nearly perfectly preserves the \hbox{$50\, \Omega$} impedance.

The materials used to build the probe were chosen to minimize heat losses. Indeed, when inserted directly in a commercial 60 L liquid helium Dewar, the probe features low helium consumption, boiling off less than 5 L a day in the absence of large heat loads. The helium consumption can be decreased by using coaxial cables with a lower thermal conductance than silver plated copper, such as silver plated cupronickel or stainless steel, at the cost of higher signal loss. For systems with strict constraints on helium consumption, additional heat sinking can be achieved by breaking the coaxial cables or wrapping them around a thermally conductive bobbin at various points along the body of the probe. The easily scalable nature of the probe allows one to add more lines while preserving optimal performance, but the helium consumption is expected to increase slightly with the addition of more coaxial lines.
\begin{figure}[t!]
\centering
\includegraphics[scale=0.42]{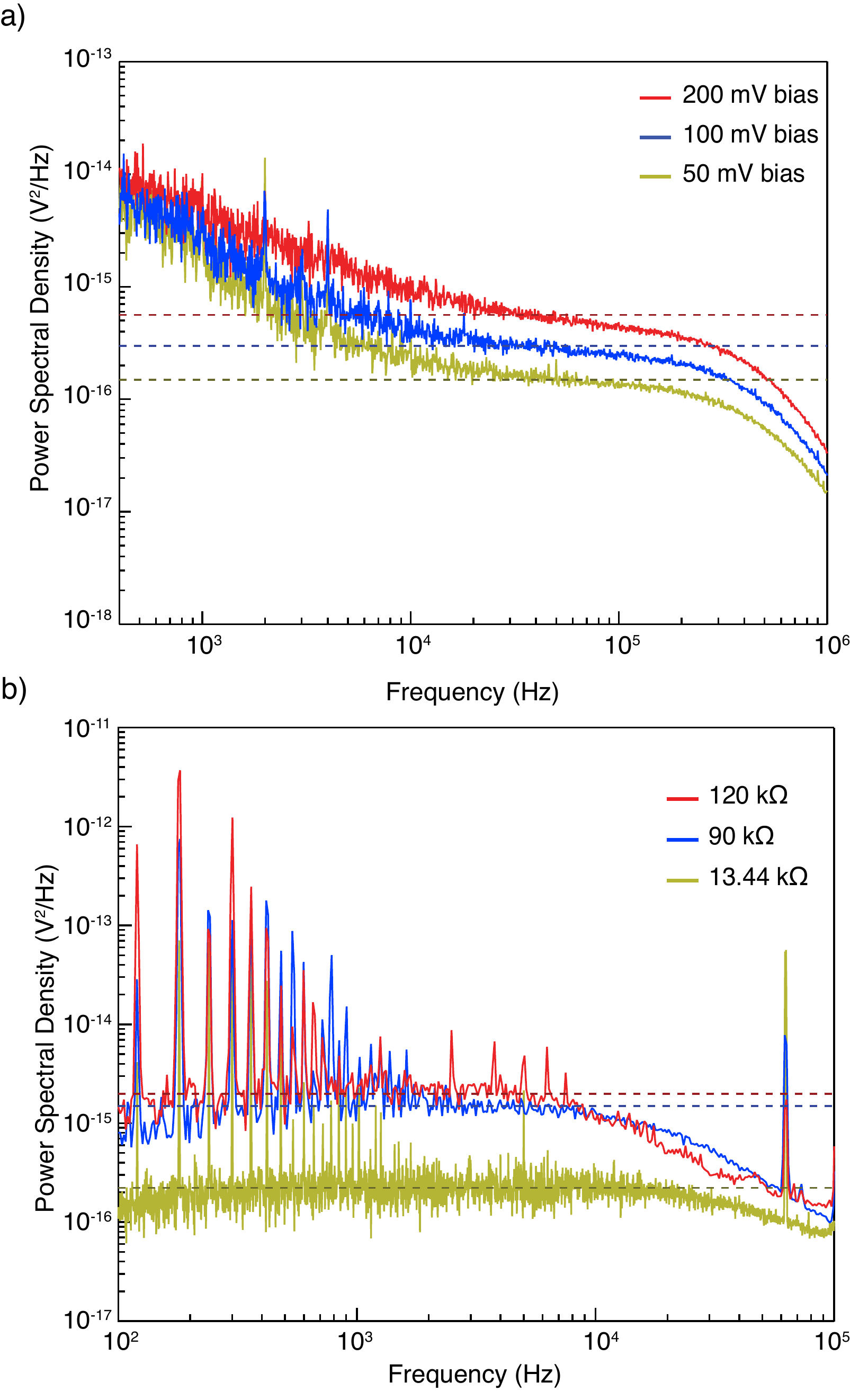}
\caption{\label{fig_noise} (a) Shot noise of an MTJ with resistance around \hbox{$9\, \text{k}\Omega$} with different bias voltages at 77 K and (b) Johnson-Nyquist noise spectra of three MTJs of different resistances measured at room temperature. Both measurements were taken on the probe with a lock-in amplifier, but (a) utilized a low temperature amplifier and (b) a room temperature amplifier (more details given in text). The expected shot noise (a), calculated using Eq. \ref{shot_eq}, and Johnson-Nyquist noise (b), calculated using Eq. \ref{Jnoise_eq}, are shown by dashed lines. The \hbox{$120\, \text{k}\Omega$} and \hbox{$90\, \text{k}\Omega$} MTJ measurements in (b) contain fewer points and thus appear to contain fewer fluctuations.}
\end{figure}

To test the probe's low-noise and low-loss properties, the shot noise of a magnetic tunnel junction was measured at three different bias voltage values, 50 mV, 100 mV, and 200 mV. The noise measurements were taken at  77 K using a low-temperature amplifier, a room temperature Femto DLPCA-200 transimpedance amplifier (TIA), and a Zurich Instruments UHFLI lock-in amplifier operating as a spectrum analyser. The lock-in amplifier measures the noise by taking the standard deviation of multiple data points collected at each frequency and normalizing to a 1 Hz bandwidth. The low-temperature amplifier we used is a SiGe heterojunction bipolar transistor (HBT) in a common emitter configuration with a high-pass, 160 Hz cut-off, RC filter between the sample and the base of the transistor to separate the transistor DC bias from the MTJ bias\cite{curry_single-shot_2019}. The HBT amplifier and MTJ are mounted in the sample holder shown in Fig. \ref{fig_probe}(c) and both fit easily into the space occupied by the microstrip in the inset of Fig. \ref{fig_s21}. There are ample coaxial connections available on the sample holder to power the low-temperature amplifier through wire bonds to the coaxial connectors, as well as to bias the MTJ (or other device under study). The resistance of the MTJ is bias and temperature dependent. At 77 K, the MTJ has resistance of \hbox{$8.98\, \text{k}\Omega$} with 200 mV bias, \hbox{$9.49\, \text{k}\Omega$} with 100 mV bias, and \hbox{$9.50\, \text{k}\Omega$} with 50 mV bias. The expected noise power spectral density is plotted as dashed lines for each bias voltage and is given by \begin{equation}
    S_V = 2eVR\coth\left(\frac{eV}{2k_BT}\right) \label{shot_eq}
\end{equation}
where $e$ is the charge of an electron, $V$ is the bias voltage on the MTJ, $R$ is the MTJ resistance, $k_B$ is Boltzmann's constant, and $T$ is the temperature\cite{he2018picotesla, lecoy_noise_1968}. In the low bias limit, when $eV\ll k_BT$, Eq. \ref{shot_eq} reduces to the Johnson-Nyquist noise equation, $S_V = 4k_BTR$, whereas in the high bias voltage limit, when $eV\gg k_BT$, it reduces to the shot noise equation, $S_V = 2eIR^2$. Figure  \ref{fig_noise}(a) shows that, for a narrow ($\sim 20-50$ kHz) frequency range, the measured shot noise is in good agreement with the expected values. The high level of low-frequency noise, the RC roll-off of the HBT-MTJ system, and the bandwidth of the TIA limit the range over which the MTJ shot noise can be measured. Various components of the low-temperature amplifier contribute to the high level of low-frequency noise that masks the MTJ shot noise for frequencies less than 10 kHz. Just as the low-frequency noise drops to levels below that of the MTJ shot noise, the signal begins to be attenuated by the RC roll-off of the HBT-MTJ system and then further by the 400 kHz bandwidth of the TIA. The ability to accurately measure MTJ shot noise at temperatures below 77 K is limited by the performance of the low-temperature amplifier. The low-frequency noise of the HBT amplifier currently employed drastically increases at 4.2 K, rendering accurate measurements of shot noise nearly impossible. Low-temperature amplifiers that remain functional at 4.2 K have been developed\cite{arakawa_cryogenic_2013, proctor_high-gain_2015, mehrpoo_cryogenic_2020, tracy_single_2016} and swapping in such an amplifier is the subject of ongoing work.

In Fig. \ref{fig_noise}(b), we show that the probe can accurately measure Johnson-Nyquist noise in excellent agreement with the expected value over a wide frequency range without the added low frequency noise of the low-temperature amplifier. In these measurements, the Johnson-Nyquist noise of three MTJs with resistances of \hbox{$13.44\, \text{k}\Omega$}, \hbox{$90\, \text{k}\Omega$}, and \hbox{$120\, \text{k}\Omega$} were measured at room temperature. The noise measurements were taken using a Stanford Research Systems SR560 low-noise preamplifier and the Zurich Instruments UHFLI lock-in amplifier. The expected noise power is plotted as dashed lines for each resistor and is given by 
\begin{equation} 
S_V = 4k_BTR. \label{Jnoise_eq}
\end{equation}
The initial deviation of the measured noise from the expected value is due to the RC roll-off of the resistor-probe system (the current coaxial lines have a capacitance of 95.1 pF/m). The noise spectra were only taken up to 100 kHz due to the limited bandwidth of the preamplifier (<1 MHz). The 60 Hz spacing of the peaks in the low-frequency regime of Fig. \ref{fig_noise}(b) indicates that it is noise introduced by the SR560 room temperature amplifier's connection to the grid and are not due to the MTJ or probe. 

While more work is necessary to improve low-temperature amplification in order to obtain accurate measurements of MTJ shot noise (i.e. spin noise in magnetic insulators), accurate measurements of Johnson-Nyquist noise, as seen in Fig. \ref{fig_noise}(b), indicate that the measurement system will be effective in measuring spin fluctuations in magnetic insulators.
Given the high sensitivity of MTJs as sensors\cite{he2018picotesla}, an accurate measurement of the MTJ Johnson-Nyquist noise can provide information on the changing resistance of the device due to the presence of an external magnetic field caused by spin fluctuations. Measuring Johnson-Nyquist noise is a well developed method of thermometry\cite{qu_johnson_2019, status_1996} and has been used to extract the value of the Boltzmann constant to high accuracy\cite{qu_improved_2017}, which indicates its reliability for extracting relevant physical parameters from a system. This ``passive'' method is preferable to measuring voltage fluctuations under bias as the MTJ resistance changes with voltage bias which would prove difficult to distinguish from changes in the sensor resistance due to spin fluctuations. Therefore, the accuracy with which our probe can detect these changes in resistance is limited by the precision with which we can measure the MTJ Johnson-Nyquist noise. Specific values of noise due to spin fluctuations, as well as the frequency of these fluctuations, are highly dependent on the sample being measured and can vary by orders of magnitude\cite{PhysRevB.99.104425}. The uncertainty and variability of the frequency of these fluctuations and magnitude of the noise motivated designing the probe to operate over a wide frequency range with minimal signal loss. Additionally, the MTJs currently at our disposal have an RC constant that limits their high frequency operation to around 1-10 MHz\cite{XiaoMTJfreq}, well within the frequency range accessible by the probe described here. The ample coaxial and DC lines available on the probe allow for the simultaneous use of two low-temperature amplifiers for cross-correlation measurements that would further increase the precision of the noise measurements.

\section{Conclusion}
The cryogenic probe design presented here is a highly modular instrument for the study of sensitive measurements, such as shot noise, that require low loss throughout a wide range of frequencies. MTJ shot noise and Johnson-Nyquist noise measurements show the probe's effectiveness in measuring small signals and S21 measurements characterize its operational ability over a wide range of frequencies and temperatures. Its simple design, as well as its highly modular nature in the interchangeability of connectors at the head of the probe and of jumper cables at the foot of the probe, allow for easy repairs and facilitates the exchange of components, which, when coupled with its low-loss capabilities, compatibility with cryogenic operation and, if necessary, high magnetic fields, makes it applicable to a wide array of measurement techniques. 

\begin{acknowledgments}
We thank Y. Zhang and A.M. Mounce for their helpful advice during the development of the probe. This work was performed, in part, at the Center for Integrated Nanotechnologies, an Office of Science User Facility operated for the U.S. Department of Energy (DOE) Office of Science. This work was supported in part by NSF grant number OMA-1936221 and DOE grant number ACOS-000R22725. E. G. was supported by the National Science Foundation Graduate Research Fellowship under Grant No. 1644760.
\end{acknowledgments}

\section*{Author Declarations}
\subsection*{Conflict of Interest}
Authors have no conflicts to disclose.

\section*{Data Availability Statement}
The data that support the findings of this study are available from the corresponding author upon reasonable request.

\section*{References}
\bibliography{noise_probe_bib}

\end{document}